# Connectopic mapping with resting-state fMRI

Koen V. Haak [* 1,2], Andre F. Marquand [1,2,4], Christian F. Beckmann [1, 2, 3]

[1]Donders Institute for Brain, Cognition and Behaviour, Centre for Cognitive Neuroimaging, Radboud University, Nijmegen, 6525EN, The Netherlands; [2]Department of Cognitive Neuroscience, Radboud University Medical Centre, Nijmegen, 6500 HB, The Netherlands; [3]Oxford Centre for Functional Magnetic Resonance Imaging of the Brain (FMRIB), University of Oxford, Oxford, OX3 9DU, United Kingdom; [4]Department of Neuroimaging, Centre for Neuroimaging Sciences, Institute of Psychiatry, King's College London, De Crespigny Park, London, SE5 8AF, United Kingdom.

**\* Corresponding author**

Koen V. Haak, PhD; Donders Institute for Brain, Cognition and Behaviour, Centre for Cognitive Neuroimaging; Radboud University, Kapittelweg 29, 6525 EN, Nijmegen, The Netherlands. E-mail: k.haak@donders.ru.nl

**Acknowledgements**

Data were provided by the Human Connectome Project, WU-Minn Consortium (Principal Investigators: D.C. Van Essen and K. Ugurbil; 1U54MH091657) funded by the 16 NIH Institutes and Centers that support the NIH Blueprint for Neuroscience Research; and by the McDonnell Center for Systems Neuroscience at Washington University. K.V.H., A.F.M. and C.F.B. gratefully acknowledge funding from the Netherlands Organisation for Scientific Research (NWO-Veni 016.171.068 to K.V.H., NWO-Vidi 016.156.415 to A.F.M. and NWO-Vidi 864-12-003 to C.F.B.) and the Wellcome Trust UK Strategic Award [098369/Z/12/Z]. C.F.B. and A.F.M. gratefully acknowledge support from the Netherlands Organisation for Scientific Research (NWO) under the Gravitation Programme Language in Interaction (grant 024.001.006).

**Highlights**

- We propose methods for mapping individualised connectopies using resting-state fMRI
- These methods include a spatial statistics approach for inference over connectopies
- The approach can tease apart overlapping connectopies that coexist within an area
- We demonstrate the methods in somatotopic and retinotopic cortex

**Abstract**

Brain regions are often topographically connected: nearby locations within one brain area connect with nearby locations in another area. Mapping these connection topographies, or 'connectopies' in short, is crucial for understanding how information is processed in the brain. Here, we propose principled, fully data-driven methods for mapping connectopies using functional magnetic resonance imaging (fMRI) data acquired at rest by combining spectral embedding of voxel-wise connectivity 'fingerprints' with a novel approach to spatial statistical inference. We applied the approach in human primary motor and visual cortex, and show that it can trace biologically plausible, overlapping connectopies in individual subjects that follow these regions' somatotopic and retinotopic maps. As a generic mechanism to perform inference over connectopies, the new spatial statistics approach enables rigorous statistical testing of hypotheses regarding the fine-grained spatial profile of functional connectivity and whether that profile is different between subjects or between experimental conditions. The combined framework offers a fundamental alternative to existing approaches to investigating functional connectivity in the brain, from voxel- or seed-pair wise characterizations of functional association, towards a full, multivariate characterization of spatial topography.

# 1. Introduction

An important open question in systems neuroscience is how the organisation of the brain in terms of its patterns of connectivity subserves its cognitive and perceptual processes. Previous work has found that brain connectivity is often topographically organised: within many brain areas, connectivity is not constant but changes gradually according to an orderly organisation wherein nearby locations connect with nearby locations elsewhere in the brain. For example, the primary cortical sensory-motor areas contain somatotopic, retinotopic and tonotopic maps, and these maps are generally continued through to higher-level sensory-motor cortex via topography-preserving cortico-cortical connections (Jbabdi et al., 2013; Kaas, 1997). Such connection topographies, which we will refer to as 'connectopies' in short, are also found in the parietal, entorhinal, parahippocampal and prefrontal cortices, as well as the basal ganglia, cerebellum and corpus callosum (Jbabdi et al., 2013; Thivierge and Marcus, 2007). The existence of connectopies is clearly at odds with the idea that the brain is organised into patches of piece-wise constant connectivity. Yet brain connectivity studies widely attempt to characterize the brain into parcels of homogeneous connectivity (Smith, 2012; Smith et al., 2013b; Power et al., 2014), while viable analysis methods for mapping these fine-grained patterns of connectivity are markedly lacking in the field (Jbabdi et al., 2013). Here, we propose a fully data-driven approach for mapping the connectopic organisation of brain areas based on functional magnetic resonance imaging (fMRI) data acquired at rest, as well as a spatial statistics approach for inference over these connectopies.

A formidable challenge to mapping connectopies concerns the fact that multiple overlapping connectopies may coexist within the same brain area of interest (Jbabdi et al., 2013; Kaas, 1997). For instance, it is well known that the connections between the early visual cortical areas are organised according to two modes of change: the distance from and the angle around the centre of the visual field (eccentricity and polar angle, respectively). For the visual brain, these overlapping modes of connectivity-change facilitate complex computations using relatively simple spatial rules and metabolically efficient short-range neural circuitry, but it poses a major yet underacknowledged obstacle for characterising the organisation of brain connectivity, possibly leading to biologically invalid results (Fig. 1). For instance, previous approaches to characterizing the topographic organization of connectivity typically involved gradually changing the position of a seed while monitoring for gradual changes in ensuing functional connectivity (Anderson et al., 2010; Buckner et al., 2011; Cauda et al., 2011; Gravel et al., 2014; Haak et al., 2013; Haak et al., 2016; Heinzle et al., 2011; Raemaekers et al., 2014; Taren et al., 2011; van den Heuvel and Hulshoff Pol, 2010), but if multiple connectopies simultaneously exist within the area of interest, these approaches will erroneously uncover their single superposition instead of revealing the true multiplicity of modes of organisation. Similar issues may also arise in the context of various connectivity-based parcellation techniques (e.g., Beckmann et al., 2005; Blumensath et al. 2013; Cohen et al., 2008; Wig et al., 2013; Wig et al., 2014).

The "moving-seed" approach is also less applicable to cortical patches with unknown topographic organization: if all goes well, the topography emerges after carefully looking at the different seed connectivity patterns linked to the seed locations, but there are many ways of traversing a region in the brain, and an exhaustive test of all possible trajectories is prohibitive. A third disadvantage is that smaller seeds produce noisier results, so the resolution at which connectopies can be reliably detected is limited. Therefore, we here draw upon the idea that all of these limitations can be overcome by reformulating the problem in terms of finding the intrinsic degrees of freedom of the high-dimensional connectivity dataset. Thus, the problem of mapping connectopies is recast as a manifold learning problem (Lee et al., 2007). Rather than moving a seed around while monitoring for gradual changes in connectivity, we compute the pair-wise similarities among the functional connectivity 'fingerprints' of all voxels within a pre-specified region of interest (ROI) and then employ manifold learning in order to find a limited set of overlapping connectopies. Because we no longer assume piece-wise constant connectivity, but attempt to characterise the topographic organization of connectivity, we also require alternative approaches to statistical inference. Therefore, we additionally propose to employ a spatial statistics approach known as trend surface analysis to perform inference over the ensuing connectopies.

In what follows, we demonstrate that the proposed approach represents a useful strategy for connectopic mapping, producing biologically valid, overlapping connectopies based on resting-state fMRI in a fully data-driven manner. We further show that the ensuing connectopies can be adequately characterised into a small number of parameters using trend surface analysis, thereby enabling statistical inference over connectopies. The combined framework offers a fundamental alternative to existing approaches to investigating functional connectivity in the brain—from voxel- or seed-pair wise characterizations of functional association and connectivity-based parcellation, towards a full multivariate characterization of spatial topographies that can be compared across subjects and experimental conditions using rigorous statistical hypothesis testing.

## 2. Methods

The proposed framework consists of three fundamental elements: connectivity fingerprinting, manifold learning, and spatial statistics for inference over connectopies (Fig. 2). We demonstrate the approach by mapping the well-known somatotopic organisation of human primary motor cortex (M1) and the retinotopic organisation of primary visual cortex (V1) using the resting-state fMRI data of 60 subjects of the WU-Minn Human Connectome Project (HCP).

### 2.1. Connectivity fingerprinting

Voxel-wise connectivity fingerprints were derived according to the following steps. (i) The fMRI time-series data from a pre-defined region-of-interest (ROI) are rearranged into a time-by-voxels matrix **A**, as are the time-series from all gray-matter voxels outside the ROI (matrix **B**). (ii) For reasons of computational tractability, the dimensionality of **B** is reduced using singular value decomposition (SVD): $\mathbf{B} = \mathbf{U}\mathbf{\Sigma}\mathbf{V}^*$. The SVD-transformed data $\widetilde{\mathbf{B}}$ can then be obtained using: $\widetilde{\mathbf{B}} = \mathbf{U}\mathbf{\Sigma}$, which transforms the data in **B** into $p = t - 1$ spatially uncorrelated components, where $t$ is the number of rows of matrix **B**. Note that here this procedure is lossless since the $p$ columns in $\widetilde{\mathbf{B}}$ explain 100% of the variance in the data in **B**. (iii) For every voxel within the ROI, its connectivity fingerprint is computed as the Pearson correlation between the voxel-wise time-series and the SVD-transformed data. This yields matrix **C**, whose rows convey correlation maps; one map for each voxel within the ROI.

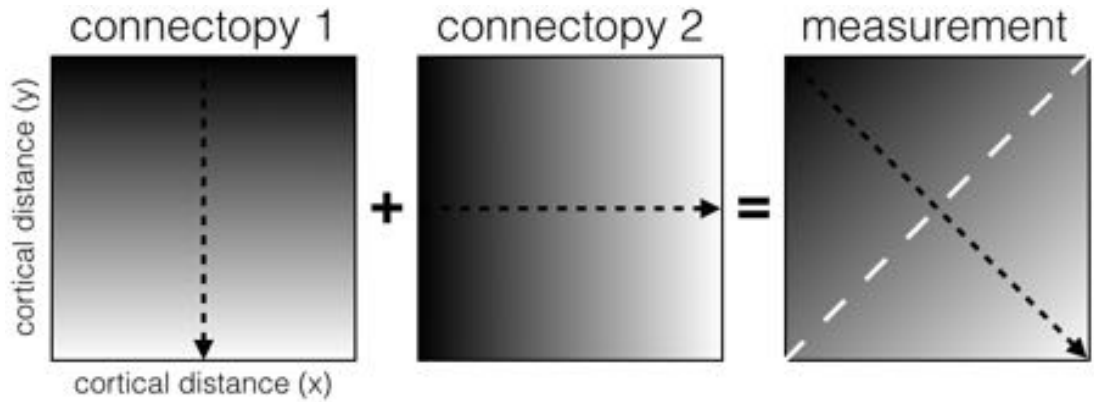

**Fig. 1.** Erroneous inference of connectivity change when overlapping connectopies co-exist within the same area of interest. Illustrated is an extreme case where one connectopy is orthogonal to a second connectopy (similar gray-tones indicate similar connectivity patterns). Unaware of the fact that these two overlapping connectopies underlie the measurements, moving a seed along the region results in the erroneous inference that the change of connectivity occurs along the diagonal. If one were to use these measurements as a basis for cortical parcellation, for instance into two sub-regions, they would lead to highly reproducible, yet erroneous subdivisions. That is, the regions would be split along the diagonal perpendicular to the diagonal direction of connectivity change (white dashed line).

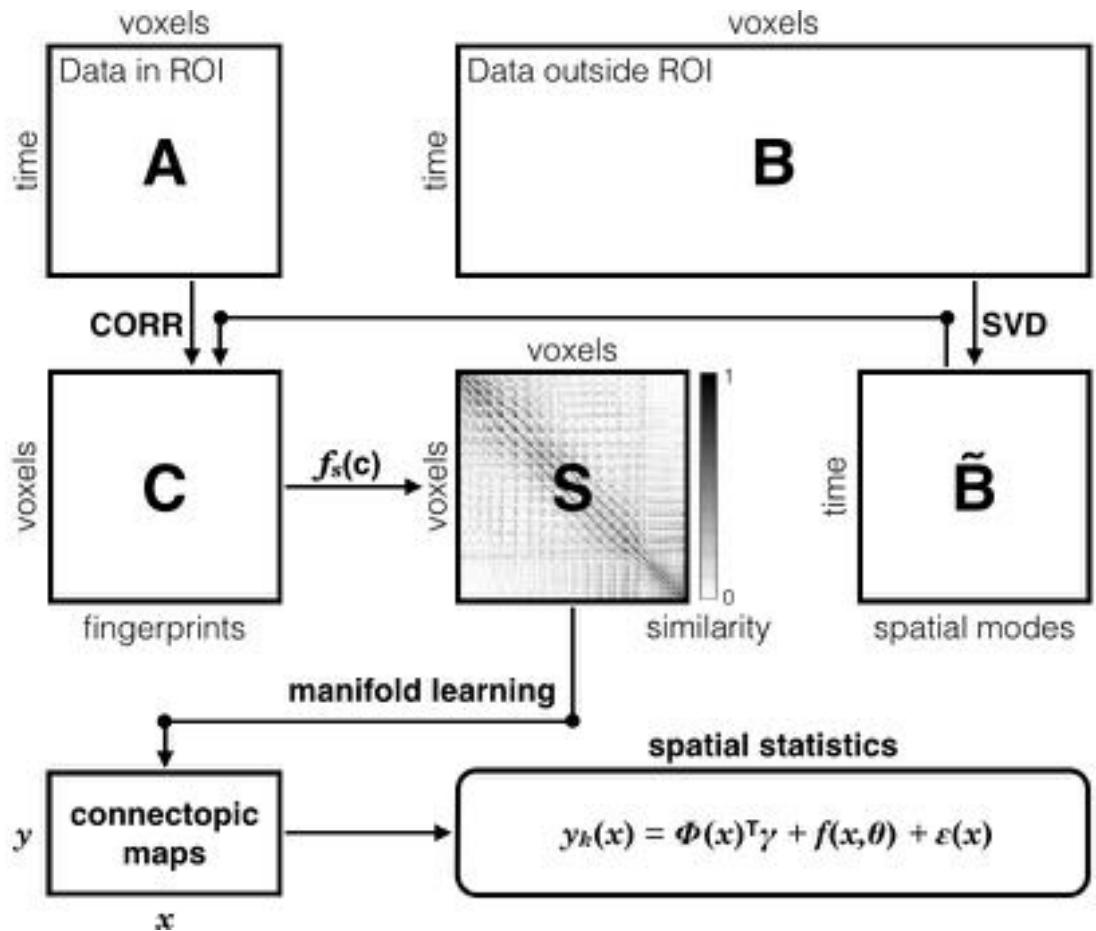

**Fig. 2.** Schematic overview of the proposed connectopic mapping framework. See Methods for details.

## 2.2. Manifold learning

We elected to employ non-linear manifold learning using the Laplacian Eigenmaps (LE) algorithm (Belkin and Niyogi, 2003), which performed well in a precursor of our framework (Navarro-Schröder et al., 2015). Known for its computational simplicity, the LE algorithm effectively represents the initial data transformation step of spectral clustering (von Luxburg, 2007), and has previously been shown to be useful for tracing changes in probabilistic white-matter tractography connectivity (Cerliani et al., 2012; Johansen-Berg et al., 2004) and connectivity-based parcellation (Craddock et al., 2012). Our implementation of the LE algorithm involves the following steps. First, the between-voxel similarity of the correlation maps is computed, yielding a matrix $\mathbf{S}$ that characterizes the within-ROI similarity of connectivity. To compute $\mathbf{S}$, we used the $\eta^2$ coefficient (Cohen et al., 2008):

$$\mathbf{S}_{\alpha,\beta} = 1 - \frac{\sum_{j=1}^{p}\left[(\mathbf{C}_{\alpha,j} - \mu_j)^2 - (\mathbf{C}_{\beta,j} - \mu_j)^2\right]}{\sum_{j=1}^{p}\left[(\mathbf{C}_{\alpha,j} - \bar{\mu})^2 - (\mathbf{C}_{\beta,j} - \bar{\mu})^2\right]} \qquad (1)$$

where $\mu_j = (\mathbf{C}_{\alpha,j} - \mathbf{C}_{\beta,j}) / 2$ and $\bar{\mu}$ is the mean of $\mu$ across all $p$ SVD-components. The $\eta^2$ coefficient represents the fraction of the variance in one connectivity profile that is accounted for by the variance in another, and ranges between 0 (entirely dissimilar) to 1 (entirely similar). In principle, other similarity measures such as the Pearson correlation or simply $\mathbf{CC'}$ could also be used. However, for these measures, negative values should first be converted to positive values so that $\mathbf{S}$ can be transformed into a graph with vertices that carry only non-negative weights (von Luxburg, 2007). Next, $\mathbf{S}$ is transformed into a connected graph represented by matrix $\mathbf{W}$:

$$\mathbf{W}_{i,j} = \begin{cases} \mathbf{S}_{i,j} & \text{if } \|\mathbf{S}_{i\cdot} - \mathbf{S}_{j\cdot}\|^2 < \varepsilon \\ 0 & \text{if } \|\mathbf{S}_{i\cdot} - \mathbf{S}_{j\cdot}\|^2 \geq \varepsilon \end{cases} \qquad (2)$$

where $\varepsilon$ is defined as the minimum value required for the graph to be connected (Cerliani et al., 2012). Note that this is done so as to meet the connected graph assumption of the next step of the analysis—or the following steps will need to be performed for each connected component separately. Ensuring a single connected component at this stage of the analysis fits with the intuition that the ensuing connectopies should cover the entire ROI instead of multiple connectopies that each might cover only a restricted portion of the ROI, depending on parameter $\varepsilon$. Given matrix $\mathbf{W}$, the graph Laplacian, matrix $\mathbf{L}$, is then set-up by computing $\mathbf{L} = \mathbf{D} - \mathbf{W}$, where $\mathbf{D}$ is a diagonal matrix with $\mathbf{D}_{i,i} = \sum \mathbf{W}_{i\cdot}$. The eigenvalues $\lambda_0 = 0 \leq \lambda_1 \leq ... \leq \lambda_k$ and eigenvectors $\{\mathbf{y}_o, \mathbf{y}_1, .., \mathbf{y}_k\}$ of the graph Laplacian are then found by solving the generalized eigenvalue problem $\mathbf{L}\mathbf{y} = \lambda \mathbf{D}\mathbf{y}$. As such, the eigenvectors $\{\mathbf{y}_1, .., \mathbf{y}_m\}$ associated with the smallest $m$ non-zero eigenvalues $\{\lambda_1, .., \lambda_m\}$ minimize $\sum_{i,j}[\mathbf{y}(i) - \mathbf{y}(j)]^2 \mathbf{W}_{i,j}$. That is, the eigenvectors convey mappings wherein voxels with similar connectivity profiles stay as close together as possible—they convey connectopies, wherein similar values represent similar connectivity patterns. The intrinsic dimensionality $m$ can be estimated using standard techniques (see e.g., Camastra and Staiano, 2016) or set in advance if the researcher is solely interested in examining the first $m$ principal modes of connectopic organisation. Given the purposes of the present paper—evaluating the proposed methods against known connectopies with known intrinsic dimensionality $m$—we used $m = 1$ for M1 and $m = 2$ for V1.

The LE algorithm represents a local non-linear approach to manifold learning (de Silva and Tenenbaum, 2003). For comparison, connectopies were also derived using a linear and a global non-linear approach. The linear approach involved a SVD of matrix $\mathbf{S}$, such that $\mathbf{S} = \mathbf{U\Sigma V^*}$ and the connectopies area given by $\mathbf{U\Sigma}$. To derive the connectopies using a global non-linear approach, we employed the Isomap algorithm (Tenenbaum et al., 2000). As such, $\mathbf{S}$ was transformed into graph $\mathbf{G}$, wherein connections exist between $k$ nearest neighbours and where $k$ was chosen as the minimum value required for the graph to be connected. Next, Dijkstra's algorithm (Dijkstra, 1959) was employed to compute the shortest path between each pair of nodes. Finally, the ensuing distance matrix $\mathbf{D}$ was submitted to multi-dimensional scaling. Specifically, a matrix $\mathbf{Q}$ was set-up by first computing $\mathbf{P} = \mathbf{D} \circ \mathbf{D}$, and then computing $\mathbf{Q} = -1/2\, \mathbf{JPJ}$, where $\mathbf{J} = \mathbf{I} - n^{-1}\mathbf{11'}$, with $n$ being the number voxels inside the ROI. From $\mathbf{Q}$, the eigenvectors $\{v_1, .., v_m\}$ associated with the $m$ largest eigenvalues were extracted, which were then transformed into the final embedding $\mathbf{Y} = \mathbf{N}\mathbf{\Lambda}^{1/2}$, where $\mathbf{N}$ is the matrix of $m$ eigenvectors, $\mathbf{\Lambda}$ is the diagonal matrix of the $m$ eigenvalues of $\mathbf{Q}$, and the columns of $\mathbf{Y}$ convey the connectopies according to Isomap.

## 2.3. Spatial Statistics

Connectopic mapping characterizes the topographic organization of connectivity in terms of a multivariate estimate (map). To enable statistical hypothesis testing on these maps, therefore, we propose a spatial statistics approach. The approach involves finding a parsimonious representation of the connectopy, governed by coefficients that can be tested either parametrically or non-parametrically or employed as features in other analyses. Finding this representation is achieved by estimating the parameters of a spatial model that describes the connectopies in terms of polynomial basis functions for the spatial trend and a Gaussian process that models more detailed spatial variation. Thus, we approximate each connectopic map using a spatial model where the value of the connectopy at spatial location $x$ is given by:

$$\boldsymbol{y}_k(\boldsymbol{x}) = \boldsymbol{\phi}(\boldsymbol{x})^T \boldsymbol{\gamma} + f(\boldsymbol{x}, \boldsymbol{\theta}) + \epsilon(\boldsymbol{x}) \tag{3}$$

Here, $\boldsymbol{y}_k$ is a connectopy, $\boldsymbol{\phi}(\boldsymbol{x})$ is a spatial basis function with coefficients $\boldsymbol{\gamma}$ that collectively describe the low-frequency spatial trend; $f(\boldsymbol{x}, \boldsymbol{\theta}) \sim \mathcal{N}(0, K(\boldsymbol{x}, \boldsymbol{x}'; \boldsymbol{\theta}))$ is a zero-mean Gaussian process that models more detailed spatial variation and $\epsilon(\boldsymbol{x}) \sim \mathcal{N}(0, \sigma_n^2)$ are spatially uncorrelated residuals. A model of this form is known as a 'trend surface model' in the spatial statistics literature (Gelfand, 2010). For the specification of the covariance function for the Gaussian process, $K(\boldsymbol{x}, \boldsymbol{x}'; \boldsymbol{\theta})$, we use a Matérn covariance function of the form:

$$K_{\text{Matérn}}(\boldsymbol{x}, \boldsymbol{x}'; \boldsymbol{\theta}) = \sigma_f^2 \frac{2^{1-\nu}}{\Gamma(\nu)} \left(\frac{\sqrt{2\nu}|x-x'|}{\ell}\right)^\nu K_\nu\left(\frac{\sqrt{2\nu}|x-x'|}{\ell}\right) \tag{4}$$

Here, $\boldsymbol{\theta} = [\sigma_f^2, \nu, \ell]^T$ where $\sigma_f$, $\nu$ and $\ell$ are respectively scaling, smoothness and length scale parameters and $K_\nu$ is a modified Bessel function (Rasmussen and Williams, 2006). This covariance function is preferred in spatial statistics over the 'squared exponential' covariance function more common in machine learning because the latter is regarded as too smooth for spatial applications (Gelfand, 2010; Wackernagel, 2003). We follow the convention in spatial statistics and fix the smoothness parameter to $\nu = 5/2$, which shows good performance in spatial applications (Wackernagel, 2003) and performed well in preliminary tests. We estimated the remaining parameters ($\boldsymbol{\gamma}$, $\sigma_n^2$, $\sigma_f^2$ and $\ell$) using nonlinear conjugate gradient optimization. See (Rasmussen and Williams, 2006) for more details.

## 2.4. Evaluation data

We evaluated our connectopic mapping approach using a data-set comprising 60 subjects of the WU-Minn Human Connectome Project (Van Essen et al., 2013), with two sessions of two 14.4-minute multi-band accelerated (TR = 0.72s) resting-state fMRI scans per individual. The 60 subjects correspond to a very initial internal HCP data-release (a subset of the Q1 release); the full list of subject numbers is available upon request. This 3T whole-brain dataset, with an isotropic spatial resolution of 2mm, is publicly available and has been pre-processed as detailed in (Smith et al., 2013a). Briefly, pre-processing steps included corrections for spatial distortions and head motion, registration to the T1w structural image, resampling to 2mm MNI space, global intensity normalisation, high-pass filtering with a cut-off at 2000s, and the FIX artefact removal procedure (Griffanti et al., 2014; Salimi-Khorshidi et al., 2014). For this work, we additionally smoothed the images and removed by regression the mean ventricular and white-matter signal from the time-series data. Spatial smoothing involved a 6mm FWHM Gaussian kernel for the analysis of M1, whereas the analysis of V1 involved different smoothing kernels for the data inside and outside V1 (3 and 6mm, respectively; the smaller smoothing kernel was used to avoid BOLD signal smearing across the upper and lower banks of the calcarine sulcus). After converting the time-series of each voxel to percent signal change by dividing by and subtracting its mean amplitude over time, the data from the two scans in each session were concatenated, resulting in two 28.8-minute functional scans per subject (one for each session day).

## 2.5. ROI definitions

Primary motor cortex (M1) was defined based on anatomical criteria using the Freesurfer toolbox. Specifically, the T1w MNI template image (1mm isotropic resolution; as provided by FSL) was subjected to Freesurfer's cortical reconstruction procedure ('recon-all') to create a number of MRI volumes wherein voxels are assigned a neuroanatomical label (e.g., left precentral gyrus). The relevant volumes were subsequently converted to the NiFTI file format, resampled to 2mm MNI space and binarized. To determine if connectopic mapping is robust to minor ROI definition inaccuracies, we further created a dilated version of the M1 ROI (FSL function "fslmaths" with option "dilM", then excluding white-matter voxels). Primary visual cortex (V1) was defined using a recent, gray-matter confined, probabilistic atlas of the retinotopic areas of the human cortical visual system (Wang et al. 2015). For each hemisphere of the brain, a mask was created of all voxels that exhibited maximum probability of being labelled as V1 (relative to the 48 retinotopic areas included in the atlas). The mask was resampled to 2mm MNI space using nearest neighbour interpolation and binarised.

### 2.6. Cross-sessions and cross-subjects reproducibility

We quantified the reproducibility of the connectopic mapping results using the intra-class correlation coefficient (*ICC* case 2,1; see Shrout and Fleiss, 1979):

$$ICC = \frac{(BMS-EMS)}{BMS+(k-1)EMS+k(JMS-EMS)/n} \quad (5)$$

where *n* is the number of 'targets' (here voxels in the ROI), *BMS* is the between targets mean square, *EMS* is the error mean square, and *JMS* is the between 'judges' mean square (here sessions or subjects). See Shrout and Fleiss (1979) for details. We quantified the cross-session reproducibility as the bootstrapped 95% confidence interval of the mean cross-session *ICC* across subjects, and the cross-subjects reproducibility as the bootstrapped 95% confidence interval of the mean cross-subjects *ICC* across pairs of subjects (this was done separately for each session day). Hence, *k* = 2 for both the cross-sessions and cross-subjects *ICC*. The bootstrap involved 1000 samples.

### 3. Results

### 3.1. Connectopic mapping at the group-level

We first describe the results at the group-level, which were obtained by computing the pair-wise similarities among the voxel-wise, whole-brain, gray-matter connectivity fingerprints in the anatomically defined ROIs (separate ROIs for each hemisphere) in each of the 60 subjects and averaging these values across them. Note that this is a valid approach for pooling data across subjects because contrary to the heterogeneous input time-series data, the similarities among the connectivity fingerprints should be broadly similar across subjects.

The human motor strip (M1) is a brain region with a well-known topographic (i.e., somatotopic) organization, well-established topographic connectivity with the motor strip in the opposing hemisphere (van den Heuvel and Hulshoff Pol, 2010) and cerebellum (Buckner et al., 2011). In line with this, the dominant, group-level connectopy ($y_1$) showed a clear correspondence with M1's somatotopic map (Fig. 3A). For further validation, we also assessed the connectopy's underlying connectivity patterns by colour-coding the voxels outside M1 according to the voxels inside M1 that they correlate the most with (Jbabdi et al., 2013). This confirmed that the connectivity patterns underlying the first connectopy can be characterized by mirror-symmetric, inter-hemispheric topographic connectivity with M1's contralateral counterpart in the opposite cerebral hemisphere (Fig. 3B) as well as its topographically organised connectivity with anterior cerebellum (Fig. 4).

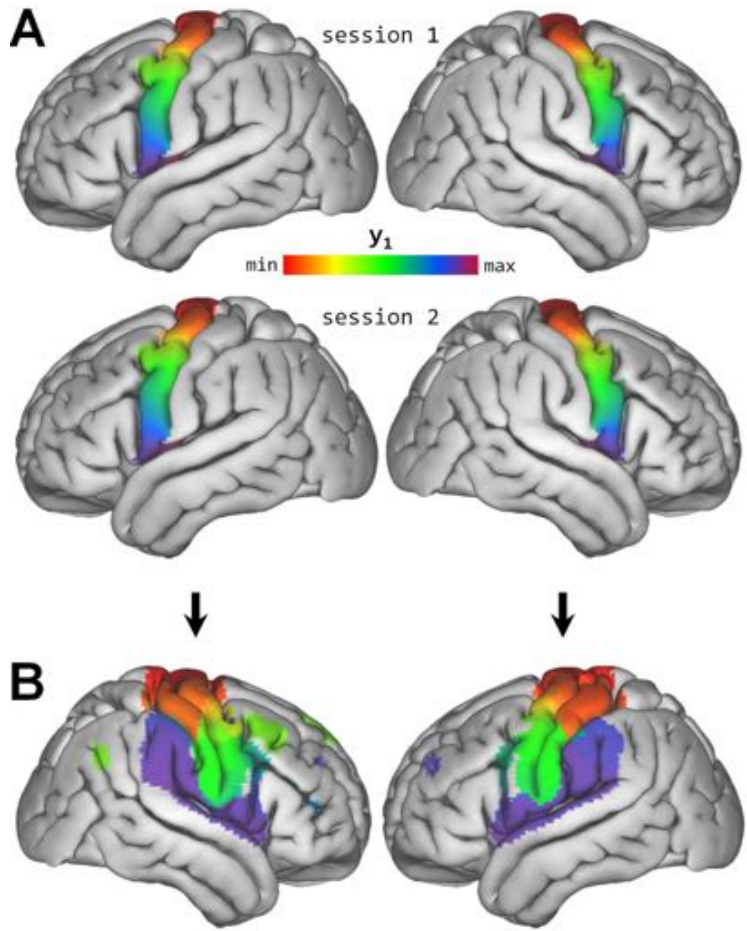

**Fig. 3.** The dominant connectopy in M1 reflects somatotopy. (**A**) The dominant group-level connectopy ($y_1$) traces M1's somatotopic map for both sessions. Connectopies were derived for left and right M1 independently. Similar colours indicate similar functional connectivity 'fingerprints'. (**B**) Projection of M1's dominant connectopy onto the opposite cerebral hemisphere. Cortical voxels are colour-coded according to the contralateral M1 voxels that they correlate the most with (max[$z$] > 10, max[$r$] > 0.2). Note how the colour-gradient in left/right M1 reappears in its contralateral counterpart in the opposite hemisphere, indicating that the two motor strips in the opposing hemispheres are topographically connected.

To determine whether connectopic mapping is also capable of tracing multiple overlapping modes of organisation, we next evaluated connectopic mapping in primary visual cortex (V1). V1 contains a retinotopic map such that distance from fixation (eccentricity) is represented along the calcarine sulcus, while the angle around fixation (polar angle) is represented orthogonal to the eccentricity map between the upper and lower banks of the calcarine sulcus (Fig. 5A). This retinotopic organisation is continued through to higher order visual cortex, and previous work has shown that it is possible to trace these retinotopic connections using resting-state fMRI, provided that the retinotopic map of V1 is known (Heinzle et al., 2011; Gravel et al., 2014; Haak et al., 2016; Glasser et al., 2016). In line with this, the dominant ($y_1$) and second-dominant ($y_2$), group-level connectopies followed clear posterior-to-anterior and superior-to-inferior trajectories, respectively (Fig. 5B). Thus, connectopic mapping can tease apart multiple overlapping modes of topographic organisation that co-exist within the same area of interest.

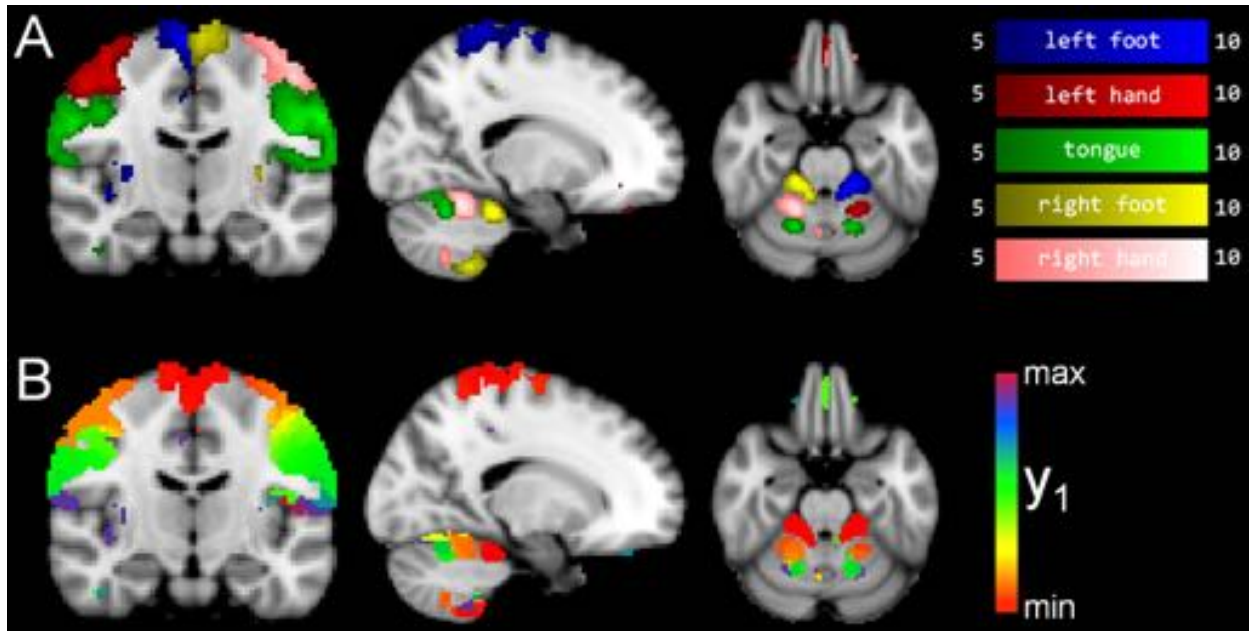

**Fig. 4.** The dominant connectopy in M1 reflects somatotopic organisation in both cerebral and cerebellar cortex. (**A**) Group-level fMRI task-activation maps ($5 < z < 10$) for the HCP motor mapping task; see (Barch et al., 2013) for details. (**B**) Projection of M1's dominant connectopy onto the portions of cerebral and cerebellar cortex that were activated during the motor mapping experiment. Cortical voxels are colour-coded according to the M1 voxels that they correlate the most with. Although the left-right asymmetry in task-activation cannot be resolved, the rfMRI-based colour-gradient follows the motor mapping results across the entire motor network, including cerebellum.

The V1 results further allow us to amplify the point made by means of the theoretical model presented in figure 1: the superposition of the two dominant modes of connectopic organisation in V1 gives rise to a highly reproducible yet biologically invalid diagonal gradient, which could in turn—when resorting to techniques that do not account for the possibility of multiple overlapping connectopies—lead to nonsensical parcellations (Fig. 5C).

### 3.2. Connectopic mapping in single subjects

We next assessed our approach' capability of mapping connectopies at the single-subject level. To this end, we applied the LE algorithm directly to the individual connectivity similarity scores, separately for each of two independent scan sessions, and computed the intra-class correlation coefficient (*ICC* case 2,1; see Shrout and Fleiss, 1979) between the ensuing connectopies. For M1, the dominant, individual connectopies showed considerable resemblance to the group result, to each other, and across the two independent sessions within the same subject (Fig. 6). They were also robust to modest ROI definition inaccuracies (Fig. 7). Table I shows that across all 60 subjects, the dominant, individual connectopies were highly similar across sessions and showed substantial correspondence across subjects. Moreover, shortening the input time-series and/or sampling every third time-point suggests that the high cross-session stability of the subject-level results generalises to scans as short as ~7.5 minutes with a TR of ~2s (Fig. 8).

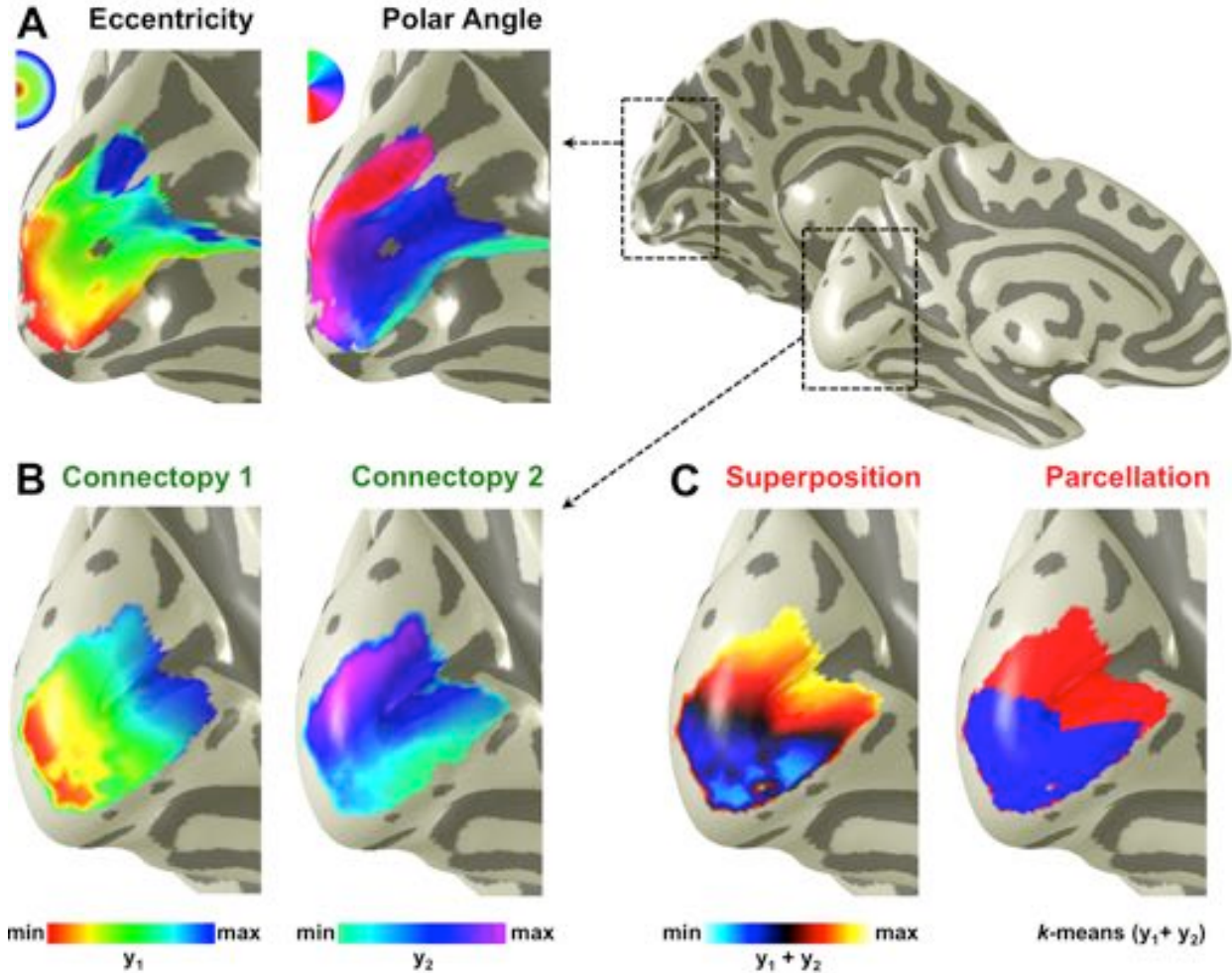

**Fig. 5.** The dominant and second-dominant connectopies in V1 reflect retinotopy. (**A**) Stimulus-based retinotopic mapping results for K.V.H. (see Haak et al., 2013 for details). Left and right panels show V1's eccentricity and polar angle maps of the right visual field in the left cerebral hemisphere. (**B**) The dominant group-level connectopy ($y_1$) follows V1's eccentricity map, while the second-dominant connectopy ($y_2$) traces V1's polar angle representation. (**C**) Using the superposition of these overlapping connectopies leads to nonsensical parcellations (see also Fig. 1).

**Table I.** Reproducibility of connectopic mapping at the single-subject level. Results are compared between sessions from the same subject and between pairs of subjects. Reported values represent the average intra-class correlation coefficient (*ICC* case 2,1) across subjects (between sessions) or subject-pairs (between subjects). Values between square brackets indicate the lower and upper bounds of the bootstrapped 95% confidence interval, respectively.

| Comparison | | | M1 ($y_1$) | V1 ($y_1$) | V1 ($y_2$) |
| --- | --- | --- | --- | --- | --- |
| Between sessions | | LH | .978 [.972, .985] | .691 [.602, .781] | .449 [.362, .536] |
| | | RH | .974 [.965, .983] | .615 [.523, .707] | .383 [.293, .472] |
| Between subjects | Session 1 | LH | .937 [.935, .939] | .603 [.589, .617] | .304 [.292, .316] |
| | | RH | .939 [.937, .942] | .540 [.524, .556] | .240 [.228, .253] |
| | Session 2 | LH | .950 [.948, .951] | .667 [.653, .680] | .270 [.257, .283] |
| | | RH | .939 [.936, .941] | .570 [.554, .585] | .259 [.245, .272] |

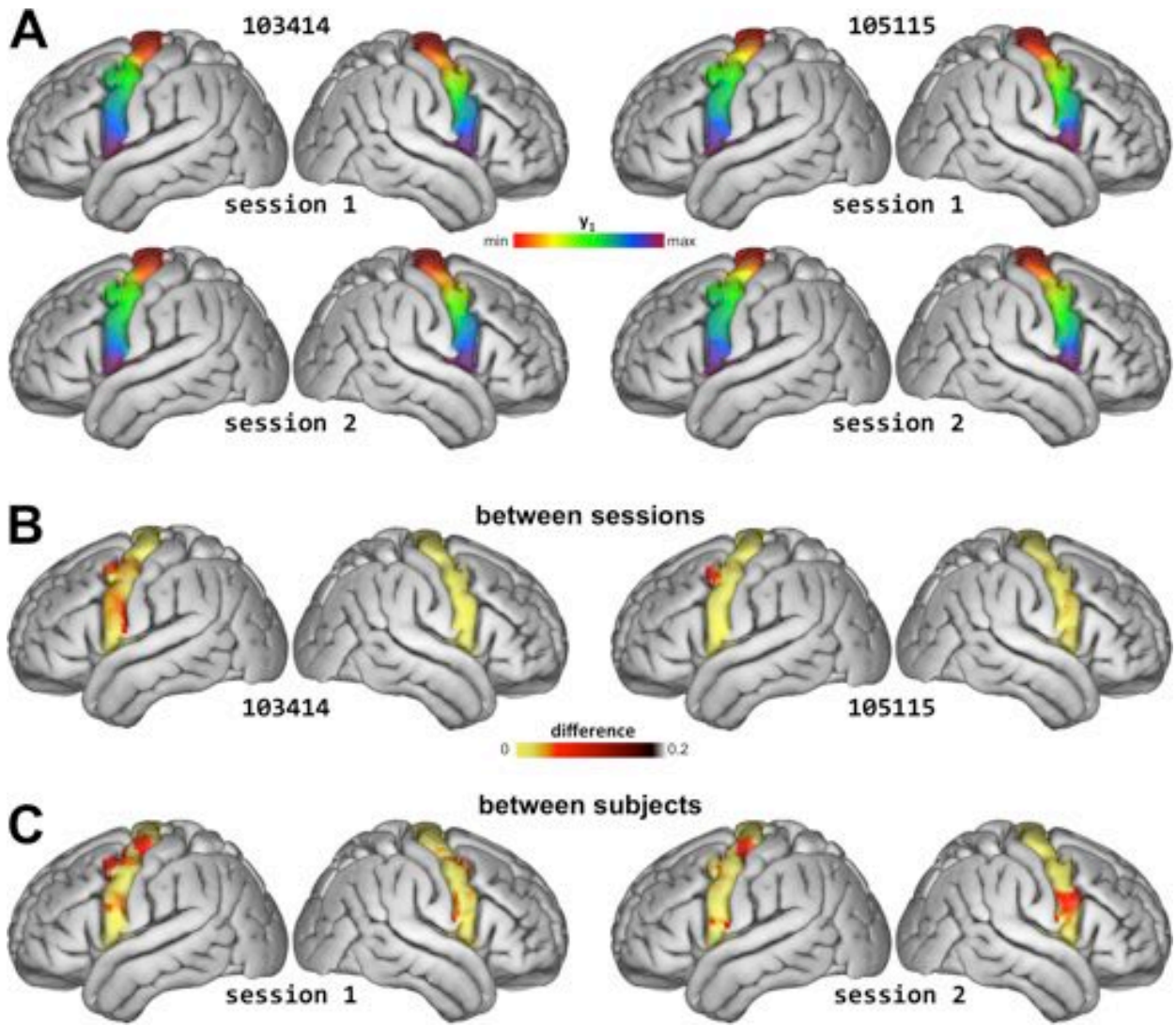

**Fig. 6.** Example subject-level results for M1. (**A**) The dominant connection-topography in M1 ($y_1$) is highly reproducible across sessions within the same subject (compare top with bottom rows). The dominant connectopy also exhibits considerable reproducibility across subjects (compare left with right panels), but to a lesser extent than across sessions within the same subject, suggesting subject-specificity. (**B**) Difference between the maps estimated from session 1 and session 2. (**C**) Difference between the maps estimated for subjects 103414 and 105115 for session 1 (left) and session 2 (right). Values represent the absolute difference after normalising $y_1$ to range between 0 and 1 for each session and subject. See table I for a quantification of these differences across all 60 subjects.

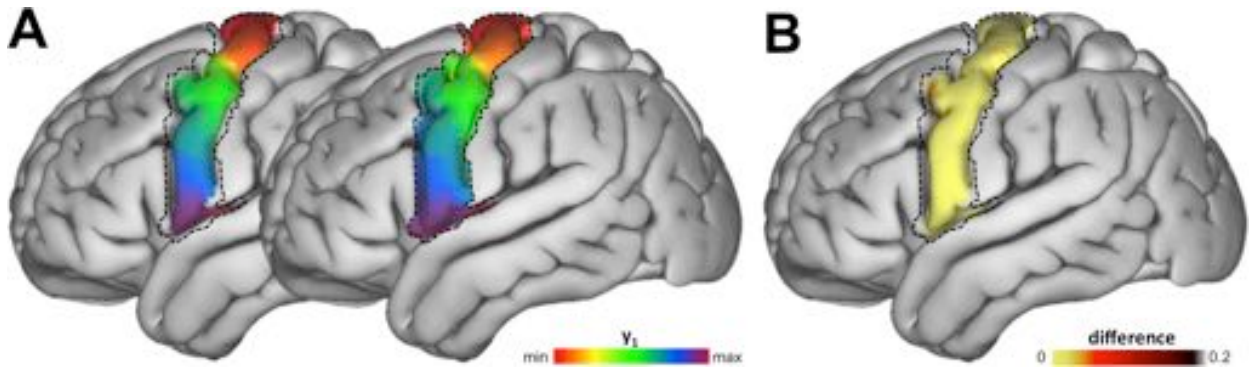

**Fig. 7.** Connectopic mapping is relatively robust to the ROI definition. (**A**) The dominant connection-topography ($y_1$) in left M1 for subject 103414 (session 1) based on a well-defined ROI (left) and a dilated version of the same ROI (right). Dashed lines indicate the contours of the dilated ROI. (**B**) Difference between these two estimates (conventions according to Figs. 6B and 6C).

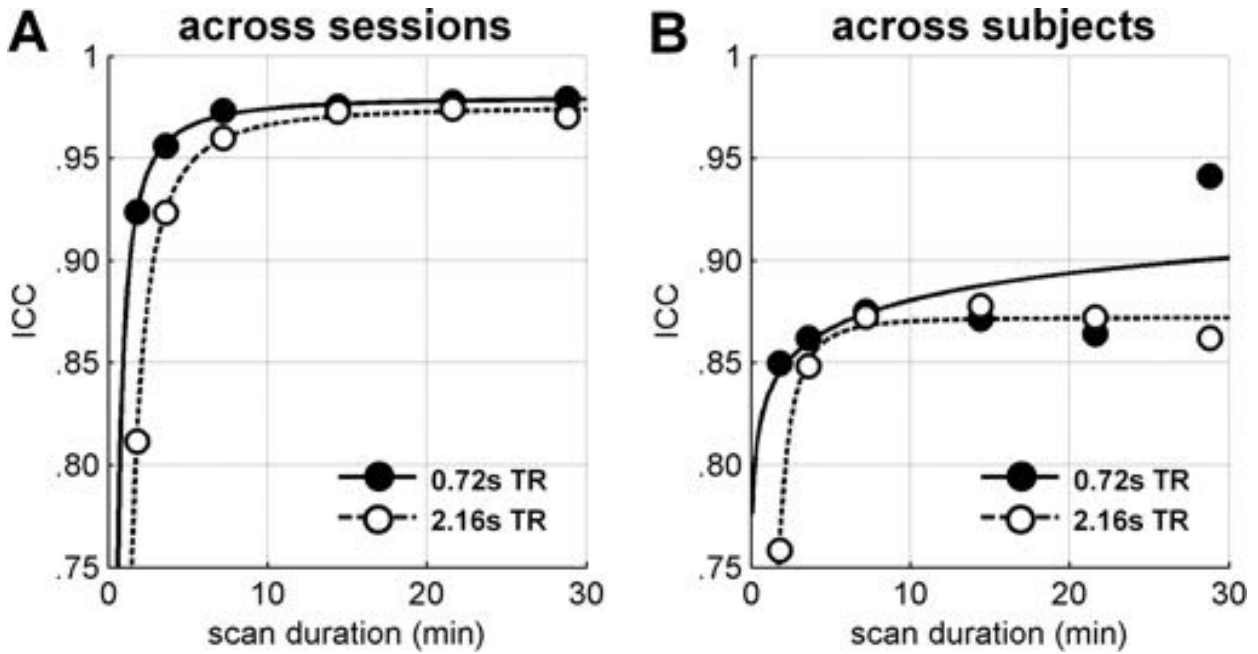

**Fig. 8**. Cross-session and cross-subjects reproducibility of the single-subject results for M1 for different scan durations and sampling frequencies. (**A**) Cross-session reproducibility results. (**B**) Cross-subjects reproducibility results. Filled and open circles indicate the average cross-session intra-class correlation coefficient (*ICC* case 2,1) for different scan-durations with the native temporal resolution (TR) of 720ms and 2.16s (by sampling every third time point), respectively. Full and dashed lines represent fits of inverse power law: $a - 1/bt^c$.

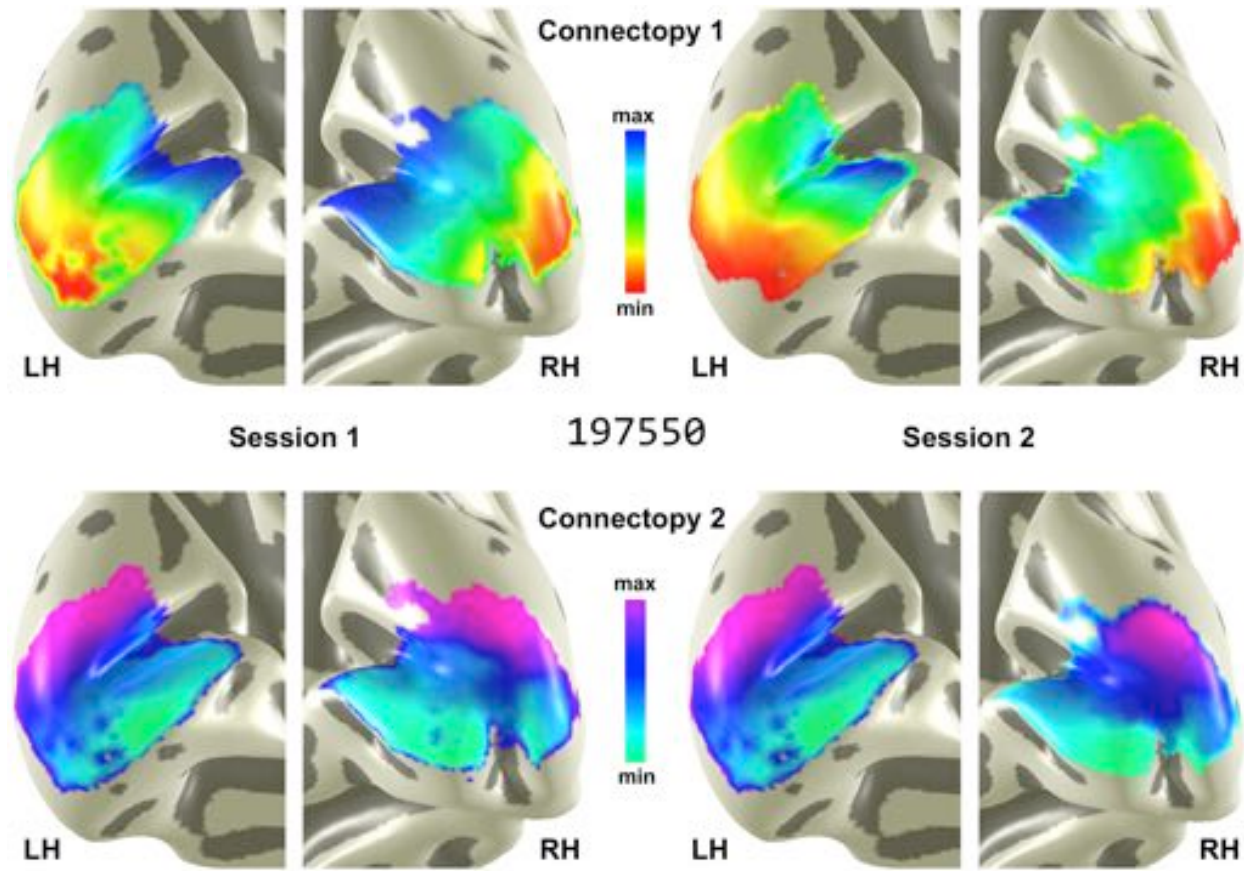

**Fig. 9.** Subject-level connectopic mapping results for V1. Conventions are according to figure 5B.

Figure 9 shows that the proposed approach can also trace V1's retinotopic map consistently across sessions in single subjects. However, where connectopic mapping could reproducibly produce plausible maps in all subjects for M1, the cross-session and cross-subjects reproducibility was much lower for V1 (Table I). Because the borders of V1 are known to be highly variable across subjects, we attribute the poorer performance for V1 to analysing the data in MNI space and consequent inaccuracies in ROI definition. In addition, the close vicinity of the upper and lower banks of the calcarine sulcus in volumetric space could have caused BOLD signal smearing across regions that are separated by a large cortical surface distance, which would particularly affect the estimates of V1's polar-angle representation ($y_2$). Future work focussing on V1 or areas with similar anatomical properties may therefore consider applying connectopic mapping in subject-native space using individualised ROIs and surface-based smoothing.

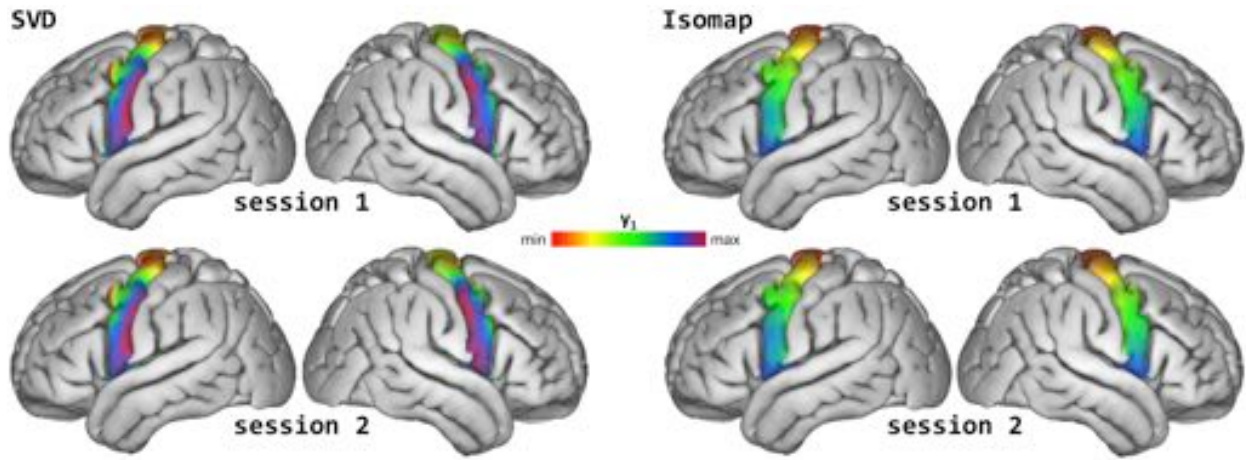

**Fig. 10.** Group-level connectopic mapping results in M1 using linear and global non-linear manifold learning. By definition, linear approaches such as SVD are not appropriate if the connectivity fingerprints sample a connectopy in a non-linear manner. In such cases, they will produce disorganised, biologically implausible connectopies compared with non-linear approaches such as LE (see Fig. 3) and Isomap (this figure, right).

### 3.3. Comparison with other manifold learning approaches

The results so far indicate that the LE algorithm represents a viable approach to mapping connectopies, but how about alternative approaches? Manifold learning algorithms fall broadly in three categories: linear approaches such as principal and independent component analysis (PCA and ICA), local non-linear approaches such as LE, and global non-linear approaches such as Isomap (de Silva and Tenenbaum, 2003). By definition, linear approaches like PCA and ICA are not able to deal with connectivity fingerprints that sample a connectopy in a non-linear manner. Figure 10 shows that such non-linear sampling does indeed occur in practice: unlike the non-linear approaches, a linear approach results in disorganised, biologically implausible connectopies. Thus, when it comes to identifying biologically plausible connectopies based on resting-state fMRI, non-linear manifold learning should be preferred over linear methods.

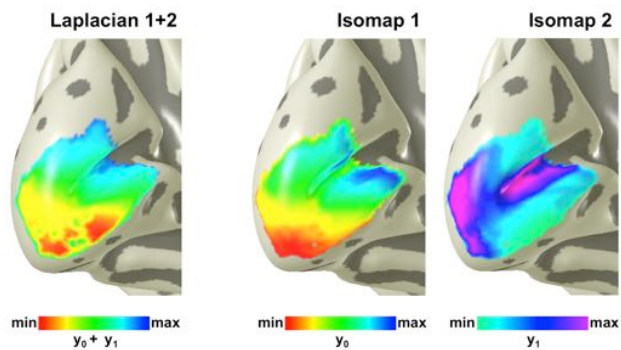

**Fig. 11.** Group-level results in V1 using Isomap. The dominant connectopy in V1 according to Isomap (middle) incorrectly corresponds to the superposition of V1's eccentricity and polar-angle organisation (left). Consequently, the second-dominant connectopy according to Isomap is also not biologically meaningful (right).

Within the domain of non-linear manifold learning algorithms, local approaches such as the LE algorithm map nearby points in the high-dimensional dataset to nearby points in the low-dimensional embedding, while global approaches such as Isomap map nearby points in the high-dimensional connectivity dataset to nearby points in the low-dimensional embedding as well as faraway points to faraway points (de Silva and Tenenbaum, 2003). The principle advantages of local approaches such as LE are computational efficiency and broader applicability, while global approaches such as Isomap tend to yield more faithful representations of the data's global geometry (de Silva and Tenenbaum, 2003). However, this focus on the global geometry appear to come at the cost of specificity at the individual level. Indeed, the reproducibility of the mappings across subjects (ICC ≈ .98) was virtually identical to the reproducibility across scan sessions within individual subjects (ICC = .98), indicating that Isomap—compared with LE—is less capable of capturing the local features of an individual's connectopic organisation. Moreover, Isomap does not appear to be capable of disentangling multiple overlapping connectopies that co-exist within the same area of interest (Fig. 11). Thus, it appears that local non-linear manifold learning approaches such as LE should be preferred over linear as well as global non-linear manifold learning.

### 3.4. Spatial statistics for inference over connectopies

Table I and figure 8B indicate that the LE-based test-retest (cross-session) reproducibility within subjects is greater than the LE-based cross-subjects reproducibility. This suggests that the proposed approach captures subject-specific features of somatotopic organization. As an example application of the proposed approach for spatial statistical inference over connectopies we tested this idea by conducting a mate-based retrieval experiment. In the context of the evaluation dataset considered here—where each subject was scanned twice—this involved aiming at retrieving the matching resting-state fMRI run for each subject. We employed a stringent (exact) matching criterion where we considered a match successful if the connectopy based on any given fMRI run achieved maximal correlation with the connectopy based on the second run from the same subject. We then computed matching accuracy as the sum of correct matches divided by the total number of fMRI runs across all subjects.

The mate-based retrieval experiment was conducted using the spatial model approximations of the connectopies. To obtain the spatial model approximations, we estimated an independent spatial model of the dominant connectopies for each session of each subject using polynomial spatial basis functions of degrees 1 through 4 and selected the optimal polynomial degree using the Bayesian information criterion (Schwarz, 1978). The optimal polynomial degree yielded fits that were nearly exact (nRMSE < $10^{-3}$) and was either two or three for all subjects, with a much smaller difference between these degrees relative to other values (e.g. first or fourth degree). To enable comparisons across subjects, we used a cubic polynomial to model the surface trend for all subjects.

We found that the spatial models allowed for correctly recognizing 62% and 53% of fMRI runs with the matching run from each subject (for the left and right ROIs, respectively), which greatly exceeds the chance level of 1/60 = 1.67% ($p < 10^{-16}$; binomial test; note that this is a much more difficult problem than standard binary classification, because chance level is 1/N, where N is the number of subjects). To establish that this identification rate ensued from differences in connectopic organisation rather than individual differences in anatomy, we repeated the mate-matching experiment asking whether a subject's spatial pattern of the mean BOLD signal over time (derived from the time-series without conversion to percent signal change) could be used to retrieve that subject's connectopy. That is, even though the functional data have been normalised to MNI space, anatomical idiosyncrasies could have introduced errors in this normalisation, which could have resulted in subject-specific spatially-varying signal amplitudes, in turn introducing artefactual connectopic features that are constant in time yet unrelated to the true patterns of functional connectivity. Indeed, the spatial pattern of the mean BOLD signal in motor cortex was near constant across sessions (nRMSE = .035, 95% CI [.031, .040] and nRMSE = .052, 95% CI [.045, .059] for the left and right hemispheres, respectively). However, crucially, it could not be used to retrieve an individual's connectopic organisation amidst those of all other individuals, as retrieval accuracy was either at or below chance level (1.67%) for both sessions and hemispheres. Thus, the proposed analysis framework reliably captures subject-specific features of connectopic organization.

## 4. Discussion

We have demonstrated that biologically meaningful, individualised connectopies can be mapped with resting-state fMRI in a principled, fully data-driven manner. This innovation builds on previous characterizations of probabilistic white matter tractography change using diffusion imaging (Cerliani et al., 2012; Johansen-Berg et al., 2004) as well as the established utility of spectral embedding for connectivity-based parcellation (CBP) (Craddock et al., 2012), and fits well with the idea to describe the functional organisation of the brain in terms of a continuous spectrum of gradual change rather than a mosaic of discrete modules or networks (Margulies et al., 2016). Our framework now allows researchers to expose the fine-grained topographic organization of a brain region's connectivity, which is discarded by CBP even though these patterns of connectivity are thought to be crucial for brain function. We have further demonstrated that our connectopic mapping approach also produces biologically valid solutions if multiple overlapping connectopies simultaneously exist within the same area under investigation, which is an important yet under acknowledged obstacle for other techniques, including various CBP approaches (e.g., Beckmann et al., 2005; Blumensath et al. 2013; Cohen et al., 2008; Wig et al., 2013; Wig et al., 2014). The presented approach does not require prior knowledge of the topographic organisation of the area under investigation, and therefore has the potential to reveal new important information about the functional organisation of the brain that might have remained inaccessible otherwise.

The shift from brain parcellation to connectopic mapping necessitates novel inference procedures and we have introduced a trend-surface analysis approach that can accurately condense the high-dimensional connectopy imaging phenotype in to a low number of trend coefficients. Trend surface analysis is a standard technique in the geosciences that is used to test hypotheses about the spatial variation of, for instance, physical geography, rainfall, temperature, political climate and so on (Gelfand, 2010). Here, we adopted trend surface analysis as a generic approach to parameterize the connectopic mapping results, thereby opening up the possibility to test hypotheses about the spatial variation of functional connectivity and whether that spatial variation is different between subjects and experimental conditions. For instance, having access to a low-parametric characterisation of connectopic organisation offers the opportunity to explicitly formulate and test anatomically relevant hypotheses beyond what can be achieved using traditional voxel- or cluster-wise testing procedures such as testing if the gradient of connectivity follows a certain anatomical orientation. Indeed, this can also be used to explicitly test the implicit assumption of CBP approaches that brain areas exhibit piece-wise constant organisation. It also affords an economical description of the spatial variation of functional connectivity for purposes such as classification. These features are likely to be of great interest to the connectivity-based cortical cartography community, as methods to perform statistical inference over the spatial layout of the brain's functional anatomy have thus far been markedly lacking.

As a demonstration of the proposed spatial statistics approach, we conducted a mate-based retrieval experiment. This yielded success rates of 55-63%, which are highly significant considering the chance-level of just 1.67%. They signify that the spatial statistical model provides an accurate yet compact description of the connectopic map estimate. The 55-63% identification rates are high considering that the analysis was confined to the human motor strip, opposed to previously work that reported a similar mate-based retrieval experiment based on whole-brain connectome data (Finn et al. 2015). Because the present characterisations were carried out in volumetric MNI space, still further improvements can be expected by employing connectopic mapping on the cortical surface in subject-native space.

The presented implementation of connectopic mapping relies on the LE algorithm for local non-linear manifold learning. The LE algorithm exists among various alternative techniques such as kernel principal component analysis (KPCA) (Scholkopf et al., 1998), isometric feature mapping (Isomap) (Tenenbaum et al., 2000), locally linear embedding (LLE) (Roweis and Saul, 2000), or the more recently introduced structural LE approach (Lewandowski et al., 2014). Each of these techniques has its own benefits and limitations and it is difficult to predict which is most fit for the task at hand. We elected the traditional LE algorithm because it has previously been found to be effective in the context of tracing changes in white-matter tractography connectivity (Cerliani et al., 2012), because of its computational simplicity, and because of its close connection to spectral clustering (Belkin and Niyogi, 2003), which is widely applied to characterize resting-state fMRI connectivity patterns. The biological plausibility of the present results demonstrate that the effectiveness of the LE algorithm also applies to capturing the fine-grained topographic structure of functional connectivity using fMRI data acquired at rest. Furthermore, a comparison with linear and global non-linear manifold learning algorithms suggests that the LE algorithm or other local non-linear manifold learning approaches should be preferred when applying connectopic mapping to datasets obtained using more conventional acquisition protocols.

To validate the biological plausibility of the ensuing connectopies across information processing hierarchies, we projected the connectopies onto the rest of cortex by colour-coding voxels outside the ROI according to the voxels inside the ROI that they correlate the most with (Jbabdi et al., 2013). This simple analysis revealed that the primary motor cortex in one hemisphere of the brain is topographically connected with the opposing hemisphere as well as anterior cerebellum, replicating previous work (Buckner et al., 2011; van den Heuvel and Hulshoff Pol, 2010). This result not only demonstrates the biological plausibility of the estimated connectopies. It also indicates that the procedure of first performing connectopic mapping in one brain region and then projecting that map onto the rest of the brain could be an effective generic approach to discover new topographically connected information processing networks. The discovery of such brain networks could be important to better understand the neural underpinnings of various perceptual and cognitive functions, e.g. by mapping such hierarchical organisation in areas beyond simple sensory cortices. In this context, one might also consider augmenting the connectopic mapping framework with more sophisticated approaches to mapping topographic organisation across information processing hierarchies such as connective field modeling (Haak et al., 2013) or a regression-based approach (Glasser et al., 2016).

A limitation of connectopic mapping is that it concerns a ROI-based analysis. ROI-based analyses are by definition dependent on the ROI definition and inaccuracies in ROI definitions could therefore affect the results. Figure 7 illustrates that the connectopic mapping results are robust to relatively modest ROI definition inaccuracies, but in principle, if a substantial portion of the ROI extends beyond the true area of interest, and if bordering areas have very different connectivity profiles, the connectopic maps will reflect the distinction between these areas rather than the true connectopic maps of the areas of interest. As with all ROI-based analyses, therefore, particular care should be taken for accurate ROI definition. Fortunately, great advances have recently been made to accurately delineate brain areas in individual subjects across the brain (Glasser et al. 2016; Laumann et al. 2015; Gordon et al. 2014; Yeo et al. 2011).

The fact that the connectopic mapping results depend to some extend on the ROI definition also affects its applicability in full brain analyses. Though it is possible in principle to apply connectopic mapping in a "ROI" that covers the entire brain (in the vein of e.g. Margulies et al. 2016), the results will reflect the overall inter-areal connectivity differences rather than the finer-grained connectivity patterns within the individual areas (because inter-areal connectivity are typically much more pronounced that intra-areal connectivity differences). Thus, this approach can be applied if the researcher is interested in capturing large-scale modes of connectivity change across many areas, but not if the interest is in the functional organisation of connectivity within individual areas. The most straightforward approach to capturing the finer-grained intra-areal connectopies in a single full brain analysis involves iteratively applying connectopic mapping to a pre-defined parcellation of the brain.

In the present work, we limited our analyses to one dominant map for M1 and two dominant maps for V1 based on the prior knowledge of topographic organization for M1 (somatotopy) and V1 (retinotopy of eccentricity and polar angle). When no such prior knowledge is available, such as in higher-order association cortex, one would have to empirically determine which of the ensuing maps are meaningful and which are not. This can be done, for instance, by plotting the residual variance $1 - R^2(\mathbf{D}_G, \mathbf{D}_S)$ against the number of maps $m$ and looking for a "knee" at which the residual variance no longer decreases significantly with added dimensions (cf. Tenenbaum et al., 2000). $\mathbf{D}_G$ is the graph distance matrix defined by the shortest path between each pair of nodes, which can be obtained by submitting matrix **W** to Dijkstra's algorithm (Dijkstra, 1959). **D**$_S$ is the matrix of Euclidean distances in the low-dimensional embedding recovered by the LE algorithm. Alternatively, for instance in cases where no clear "knee" can be observed, one could resort to one of several more sophisticated approaches to estimate the intrinsic dimensionality of the data (for a recent review see Camastra and Staiano, 2016).

Previous work on probabilistic tractography data acquired with diffusion imaging has also applied manifold learning in an attempt to find anatomical gradients of connectivity (Cerliani et al., 2012; Johansen-Berg et al., 2004). However, white-matter tractography is limited in the ability to trace the precise site-to-site connections required for mapping connectopies (Jbabdi et al., 2013; Jbabdi and Johansen-Berg, 2011). Resting-state fMRI measures connectivity directly at each voxel and these measurements are one of function. The ability to map and perform inference over resting-state fMRI connectopies thus opens up a wide range of novel research opportunities. For instance, our framework can be employed to test for the presence of topographic maps in association cortex, which was not possible in the past because the existing stimulus- and task-based approaches required detailed knowledge of the information that is represented in these areas; to reveal the topographic organisation of the visual areas we know we are to scale or rotate a visual stimulus, but it is less clear along what dimensions the stimulus or task should change to reveal topographic maps in association cortex. Testing for resting-state connectopies instead offers a unique new angle to investigating these regions' functional organisation and may thus drive future studies for better understanding the fundamental nature of the regional computations. Likewise, connectopic mapping could provide a translational avenue to patients with impairments that preclude stimulus-driven or task-based experiments. Because topographic maps are widely thought to be crucial to healthy brain function, connectopic mapping also holds great promise for developing more sensitive markers of disease, advancing both cognitive and clinical imaging neuroscience.